\documentclass[11pt]{article} 
\usepackage{graphicx}
\usepackage{hyperref}
\usepackage{amssymb}

\usepackage{authblk}

\date{}

\title{Analysis of vertex-contained high energy neutrino events for the KM3NeT/ARCA detector.}

\author{K. Pikounis\thanks{pikounis@inp.demokritos.gr} }
\author{E. Tzamariudaki on behalf of the KM3NeT collaboration}
\affil{Instistute of Nuclear and Particle Physics, N.C.S.R. Demokritos, Athens, Greece}

\begin{document} 

\maketitle 

\begin{abstract}
KM3NeT is a research infrastructure housing the next generation neutrino detectors in the depths of the Mediterranean Sea. The ARCA detector, which is currently under construction, is optimized for searches for neutrinos from astrophysical sources as well as measurements of the diffuse astrophysical flux. The unambiguous detection of neutrinos of extraterrestrial origin by IceCube has led to the first measurement of a high energy astrophysical neutrino flux. The properties of sea water allow for a measurement of the neutrino direction with an excellent angular resolution for both track and cascade events. Here a method to differentiate track and shower events and a method to reject the atmospheric muon background from starting track-like events are combined in one analysis. The analysis on the discovery potential of KM3NeT/ARCA for a diffuse astrophysical neutrino flux using events with the reconstructed vertex inside the detector volume will be presented.
\end{abstract}

\section{Introduction}
KM3NeT~\cite{LOI} is a research infrastructure housing the next generation neutrino detectors in the depths of the Mediterranean Sea. The  main science objectives of the ARCA (Astroparticle Research with Cosmics in the Abyss) telescope are the detection of neutrinos from astrophysical sources and the measurement of the diffuse astrophysical flux. ARCA is currently under construction at a depth of 3500m, approximately 80km off-shore Portopalo di Capo Passero in Sicily. When completed, ARCA will consist of two building blocks of 115 vertical detection units, each hosting 18 Digital Optical Modules (DOMs), providing an instrumented volume of about 1 km${^3}$.The DOM is a high pressure resistant glass sphere containing 31 3-inch PhotoMultipliers (PMTs) and the related electronics. \par
The unambiguous detection of neutrinos of extraterrestrial origin by IceCube~\cite{Ice_HESE} has led to the first measurement of a high energy astrophysical neutrino flux. Atmospheric muons, which constitute the most prominent and high-rate background to the astrophysical neutrino signal, can be suppressed either by selecting upgoing events, or by requiring that the reconstructed vertex lies inside the instrumented volume. The properties of sea water combined with the cutting-edge technology used for the multi-PMT Digital Optical Modules in the KM3NeT detectors, allow for a measurement of the neutrino direction with an excellent angular resolution for both track and cascade events. Taking advantage of this accuracy, tools to differentiate track-like from shower-like events and to reject incoming track events to the ARCA detector were developed.

\section{High Energy Starting Tracks (HEST) Analysis}

\begin{figure}[ht]
  \centering
  \includegraphics[width=0.49\textwidth]{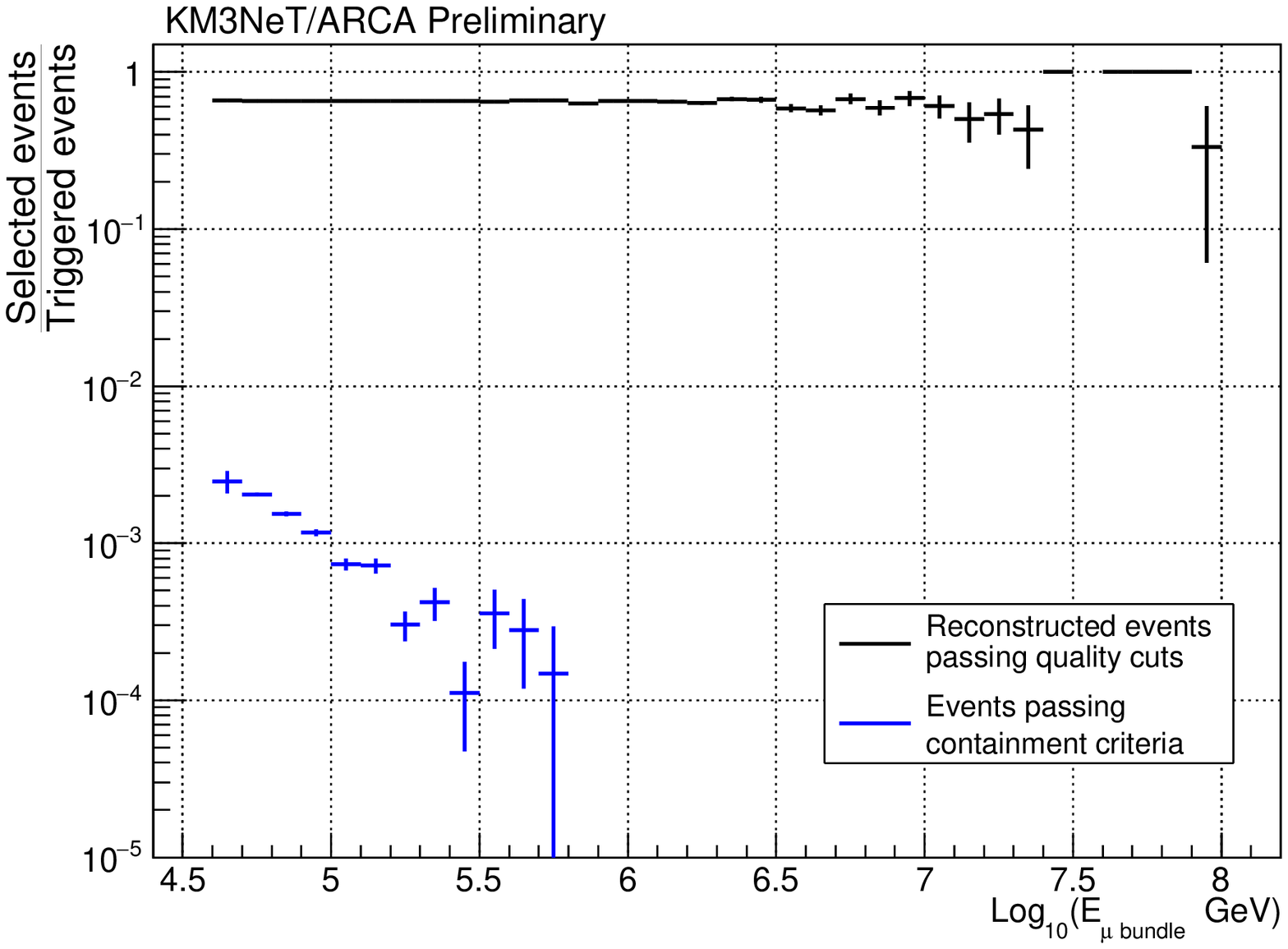}
  \includegraphics[width=0.49\textwidth]{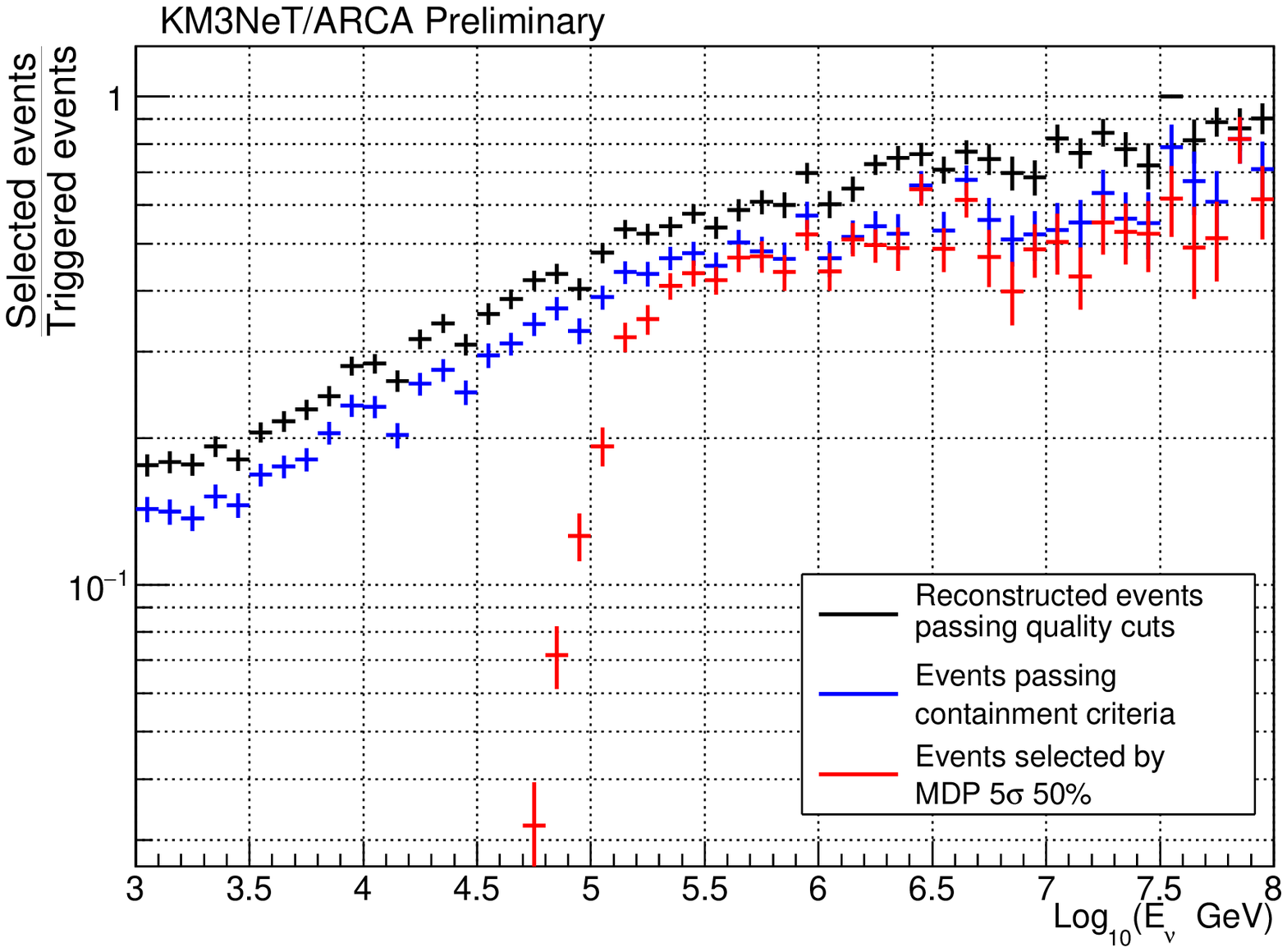}
  \caption{Ratio of the number of events surviving each of the selection requirements over the number of triggered events. Left:  Sample of atmospheric muon events with respect to E$_{bundle}$. Right: $\nu_{\mu}$ and $\overline{\nu_{\mu}}$ with respect to the true E$_{\nu}$.  }
  \label{f:HEST_eff}
\end{figure}

A sample of HEST events can be obtained using the methodology designed to reject incoming track-like events to ARCA. At first, well reconstructed track events are selected by applying quality criteria on the track reconstruction output. In addition, the reconstructed interaction vertex is required to lie inside a volume slightly smaller than the actual volume of the detector. The containment condition results in the rejection of the vast majority of incoming track events as illustrated in Figure~\ref{f:HEST_eff} (left), while a high efficiency is retained for true HEST events as shown in Figure~\ref{f:HEST_eff} (right) (in blue). The final identification of the HEST events is performed by a Boosted Decision Tree (BDT) using ten event-based variables. \par

\begin{figure}[ht]
  \centering
  \includegraphics[width=0.75\textwidth]{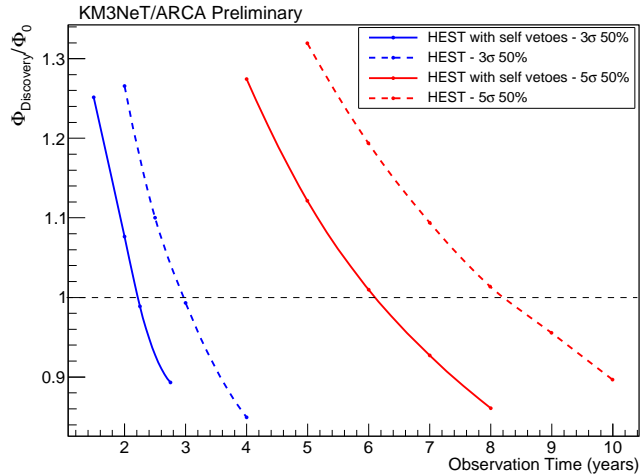}
  \caption{Ratio between the discovery flux normalization factor and the expected neutrino flux normalization $\Phi_0$ , for 3$\sigma$ (blue) and 5$\sigma$ (red) with 50\% probability for the full ARCA detector, as a function of the observation time in years. Solid lines: both atmospheric neutrino and CORSIKA MC samples used for the atmospheric neutrino background, dashed lines: only atmospheric neutrino MC sample used. }
  \label{f:HEST_Disc_pot}
\end{figure}

A discovery potential is estimated using only HEST events with the Model Discovery Potential~\cite{MDP} (MDP) minimization technique adopting the cut and count approach~\cite{LOI}. The astrophysical flux used was:
\begin{equation}
\Phi_{astro} = \Phi_0\cdot(E/{100 TeV})^{-\gamma}
\end{equation}
with $\Phi_0 = 2.3 \cdot 10^{-18} GeV^{-1}cm^{-2}s^{-1}sr^{-1}$ for each neutrino flavor and $\gamma = 2.5$ .
The Honda flux~\cite{Honda} with a prompt component as calculated by Enberg~\cite{Enberg} and a correction taking into account the “knee” of the cosmic ray spectrum is assumed for the background of atmospheric neutrinos. Atmospheric muon bundles were simulated with MUPAGE~\cite{MUPAGE} code. As illustrated in Figure~\ref{f:HEST_Disc_pot} (dashed lines), a 3$\sigma$ and a 5$\sigma$ discovery of the astrophysical flux with a 50\% probability is expected to be achieved by ARCA in 3 and 8 years of observation time, respectively. The efficiency of all selection criteria applied is illustrated in Figure~\ref{f:HEST_eff} (right) for truly contained $\nu_{\mu}$ and $\overline{\nu_{\mu}}$ events. The MDP minimization yields an energy cut E$_{reco}$ $\geq$ 10$^{5.0}$ GeV for the final event sample. \par
In order to evaluate the possibility of vetoing atmospheric neutrinos which are accompanied by muons produced at the same atmospheric shower, the analysis was repeated using as atmospheric neutrino events coming from above a sample simulated by CORSIKA~\cite{CORSIKA}. Practically all atmospheric neutrinos accompanied by muons were rejected resulting in a reduction of the atmospheric neutrino background by 32\%, and a corresponding decrease of the observation time for a 3$\sigma$ and a 5$\sigma$ discovery as illustrated in Figure~\ref{f:HEST_Disc_pot} (solid lines).

\section{High Energy Starting Events (HESE) Analysis}
\begin{figure}[ht]
  \centering
  \includegraphics[width=0.49\textwidth]{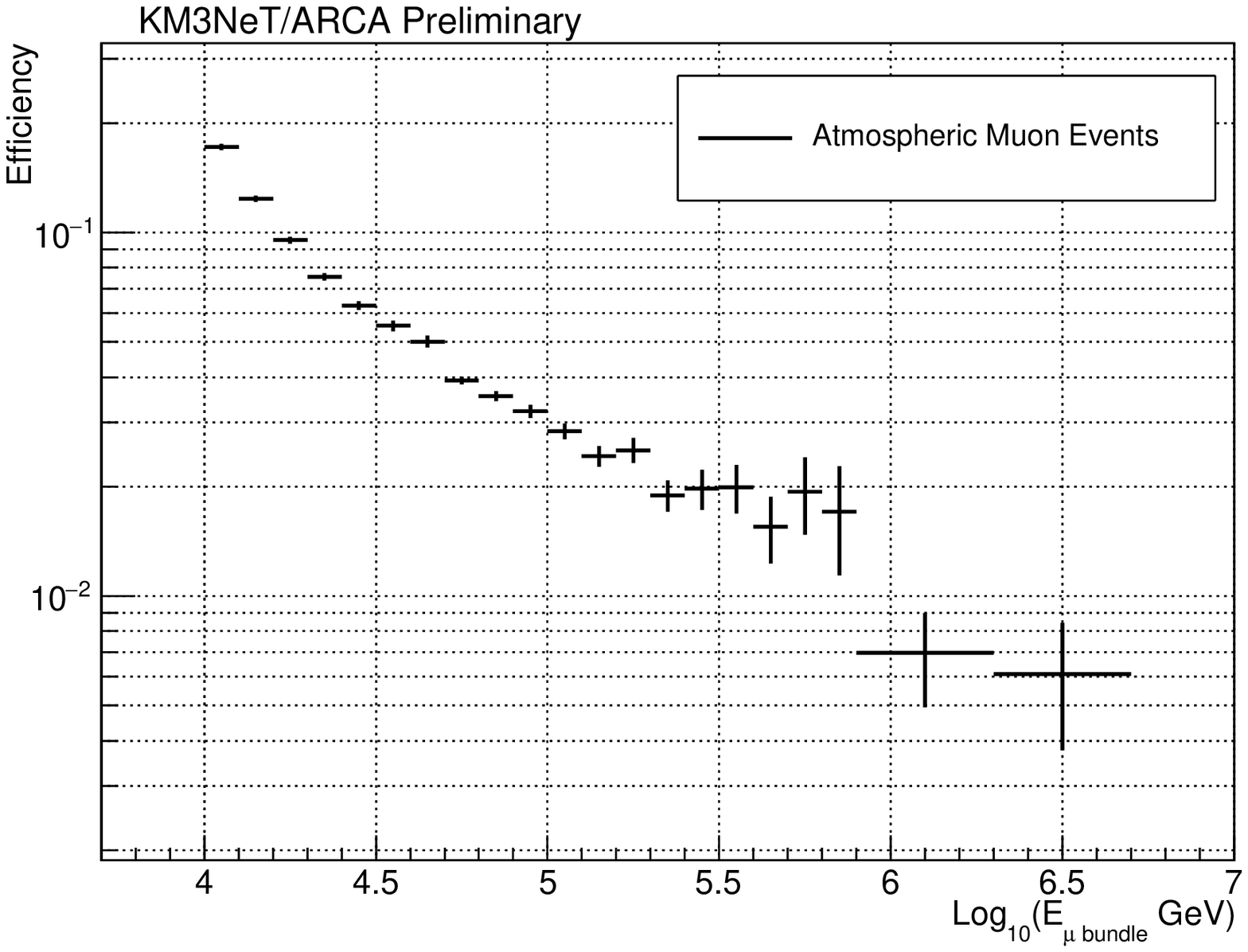}
  \includegraphics[width=0.49\textwidth]{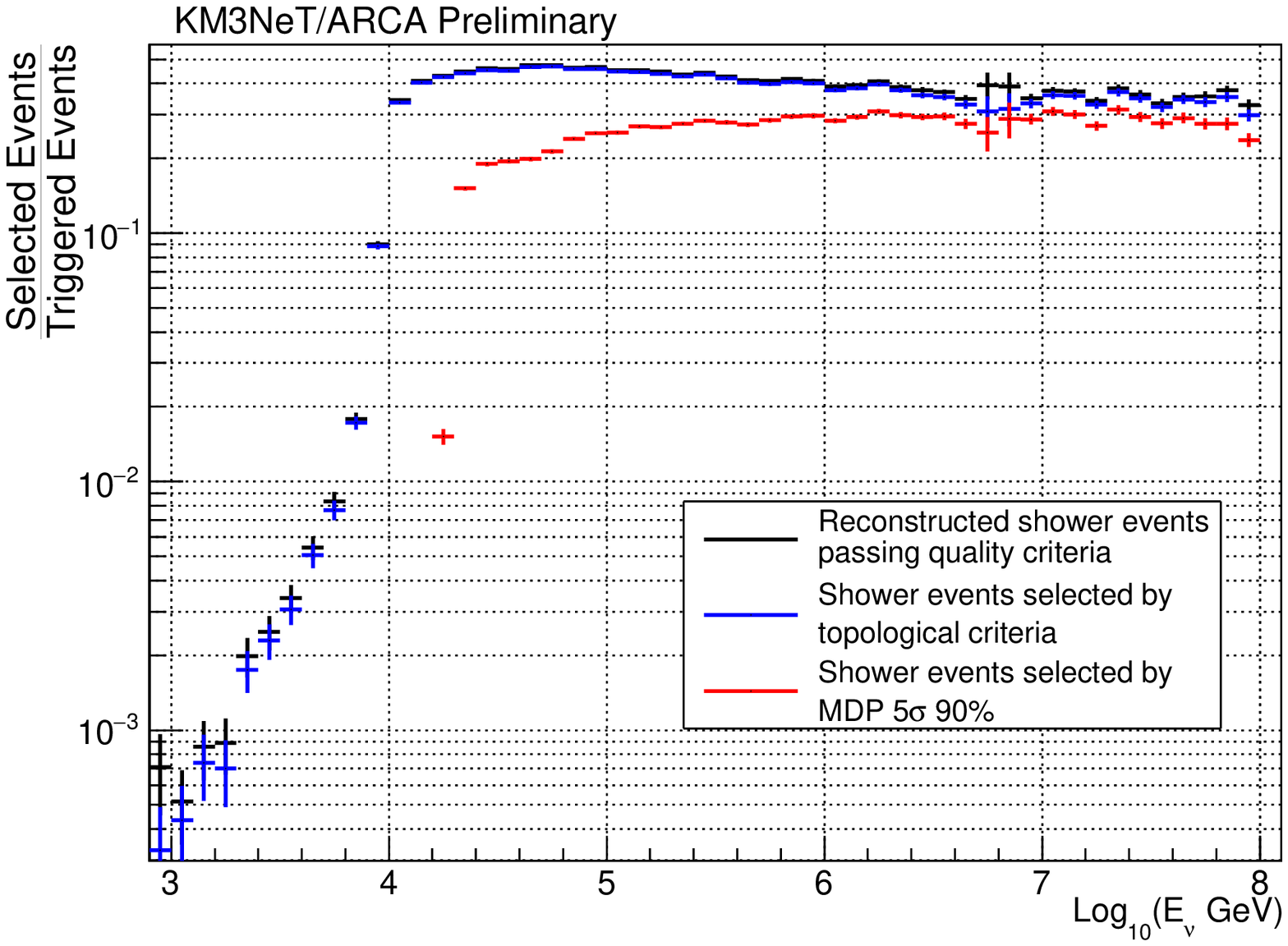}
  \caption{Left: Efficiency, ratio of the selected events by the topological criteria over all events characterized as well reconstructed, for atmospheric muon events. Right: Ratio of number of events surviving each cut over the number of triggered for all true showers events with respect to the true E$_{\nu}$.}
  \label{f:HESE_eff}
\end{figure}

A tool based on a BDT was designed to differentiate between shower-like and track-like events and select the former. Firstly, well reconstructed shower events, with a reconstructed vertex inside the volume of the detector, are selected by applying quality criteria on the shower reconstruction output. A series of topological selection criteria are further applied in order to reject track-like events. The rejection power for track events is illustrated in Figure~\ref{f:HESE_eff} (left) where atmospheric muon events are used as a pure track sample. The selection efficiency for real shower events is shown in Figure~\ref{f:HESE_eff} right. For the final identification of shower-like events, a BDT with twelve event-based variables is employed. \par

\begin{figure}[ht]
  \centering
  \includegraphics[width=0.75\textwidth]{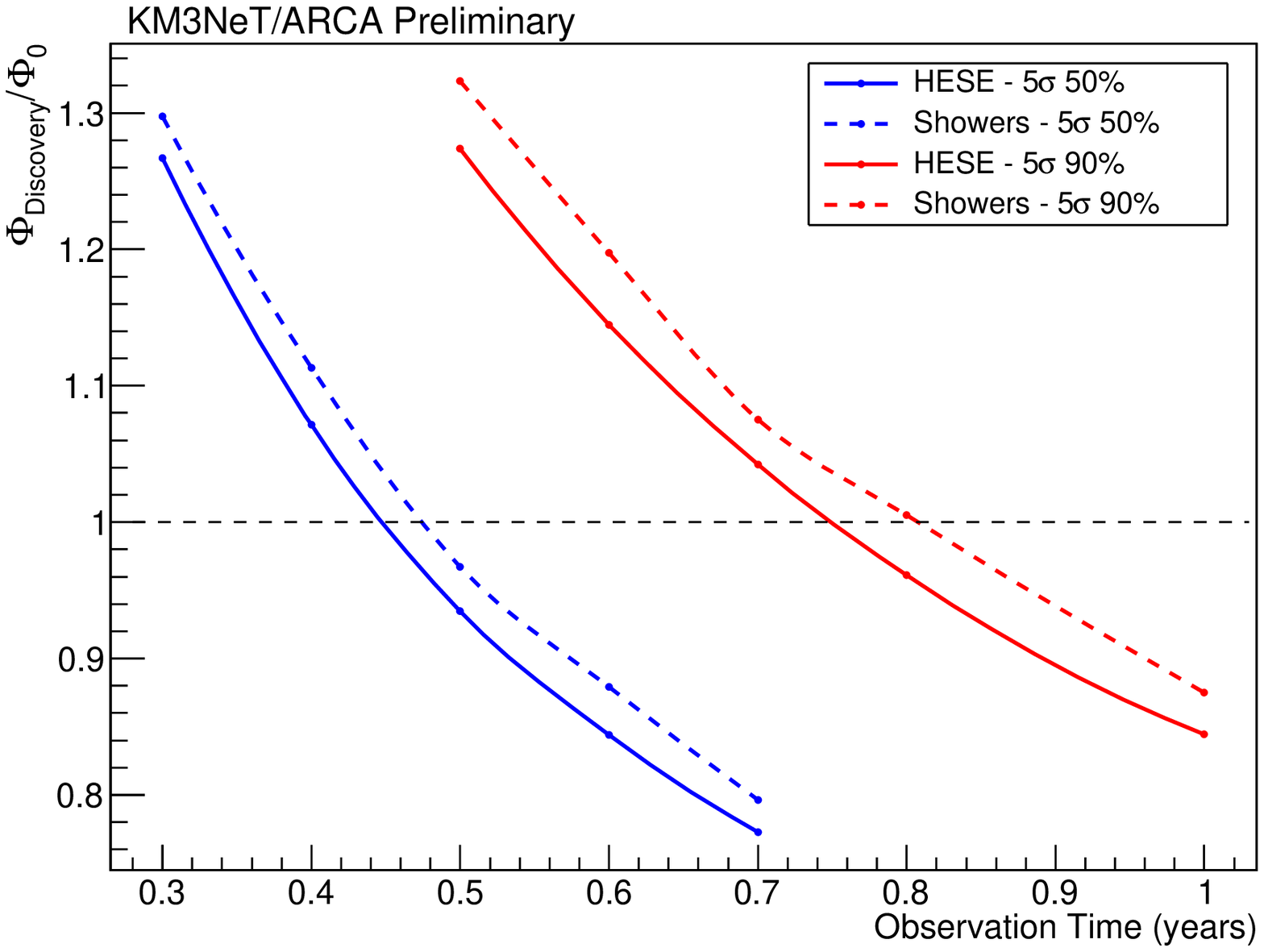} 
  \caption{Ratio between the discovery flux normalization factor and the expected neutrino flux normalization $\Phi_0$ , for 5$\sigma$ with 50\% probability (blue) and 90\% probability (red) for the full ARCA detector, as a function of the observation time in years. Solid lines: HESE analysis, dashed lines: only the shower sample.}
  \label{f:HESE_Disc_pot}
\end{figure}

The selected sample of events is used to estimate the discovery potential with the cut and count approach. The astrophysical and atmospheric neutrino fluxes used are the same as in the HEST analysis. Atmospheric muon bundles simulated with MUPAGE were also used. A 5$\sigma$ discovery of the IceCube astrophysical flux~\cite{Ice_Flux} can be achieved in the cascade channel with a probability of 50\% and 90\% after 0.5 and 0.8 years of ARCA  data taking, as shown in Figure~\ref{f:HESE_Disc_pot} (dashed lines).\par

To further improve the ARCA discovery potential, the shower sample is combined with the HEST sample to obtain a sample of High Energy Starting Events (HESE). A 5$\sigma$ discovery of the IceCube astrophysical flux can be achieved with a probability of 50\% in less than 0.5 years and 90\% in less than 0.8 years of ARCA observation, as shown in Figure~\ref{f:HESE_Disc_pot} (solid lines). In the final sample of shower-like events, as obtained after the application of the cuts provided by the MDP minimization for 5$\sigma$ with 90\% probability, more than 85\% of the events are real shower events. The contamination of atmospheric muons is less than 5\%, while CC $\nu_{\mu}$ and $\overline{\nu_{\mu}}$ interactions at the edge of the detector with the muon leaving the detector, dominate the rest 10\% of misclassified events, as they leave a shower-like signature. The percentage of correctly identified track events in the HEST samples exceeds 92\%.

\section{Conclusion}
Two BDT based tools were developed to identify the HEST and the contained shower-like events. These tools were combined to perform a HESE analysis for the KM3NeT/ARCA detector. The effect of vetoing atmospheric neutrinos accompanied by muons created at the same atmospheric shower was explored using the HEST sample, where the contribution of atmospheric neutrino is more prominent. The total number of atmospheric neutrino background events can be reduced by 32\% by exploiting self vetoes. Without including the self-veto effect for atmospheric neutrinos, the ARCA detector is expected to make a 5$\sigma$ discovery of the IceCube flux with 50\% and 90\% probability in less than 0.5 and 0.8 years, respectively. \par

\end{document}